\documentclass[%
 reprint,
superscriptaddress,
showpacs,
 amsmath,amssymb,
 aps,
floatfix,
]{revtex4-1}

\usepackage{dcolumn}
\usepackage{bm}

\usepackage[dvipdfmx]{graphicx,color}

\usepackage{ulem} 

\begin{document}


\title{Photoinduced absorptions inside the Mott gap in the two-dimensional extended Hubbard model}
\author{Kazuya Shinjo}
\author{Takami Tohyama}
\affiliation{Department of Applied Physics, Tokyo University of Science, Tokyo 125-8585, Japan}

\date{\today}
             

\begin{abstract}
We theoretically investigate pump-probe optical responses in the two-dimensional extended Hubbard model describing cuprates by using a time-dependent Lanczos method. At half filling, pumping generates photoinduced absorptions inside the Mott gap. A part of low-energy absorptions is attributed to the independent propagation of photoinduced holons and doublons. The spectral weight just below the Mott gap increases with decreasing the on-site Coulomb interaction $U$. We find that the next-nearest-neighbor Coulomb interaction $V_1$ enhances this $U$ dependence, indicating the presence of biexcitonic contributions formed by two holon-doublon pairs. Photo-pumping in hole-doped systems also induces spectral weights below remnant Mott-gap excitations, being consistent with recent experiments. The induced weights are less sensitive to $V_1$ and may be related to the formation of a biexcitonic state in the presence of hole carriers.

\end{abstract}
\maketitle


\section{Introduction}

Pump-probe spectroscopy is a good tool to characterize electronic states in strongly correlated electron systems. One of the examples is pump-probe optical measurements in the two-dimensional Mott insulators, La$_2$CuO$_4$ and Nd$_2$CuO$_4$. Photoinduced midgap excitations inside the Mott gap have been reported just after pumping~\cite{Okamoto11}. The excitations at midinfrared region have been assigned to those due to photoinduced holons and doublons through the comparison with absorption spectra in La$_{2-x}$Sr$_x$CuO$_4$ and Nd$_{2-x}$Ce$_2$CuO$_4$. On the other hand, photoinduced excitations just below the Mott gap have been attributed to thermalization effects, consistent with thermal broadening of absorption spectra~\cite{Okamoto11, Novelli14}. Even for hole-doped cuprates, similar pump-probe experiments have been done and the emergence of spectral weight just below Mott-gap excitations has been reported~\cite{Peli17}. However, there are only a few theoretical investigations on such pump-probe experiments in cuprates~\cite{Filippis12, Novelli14} and related works~\cite{Zala14,Eckstein16,Denis17}.

The one-band Hubbard model with on-site Coulomb interaction is commonly used for describing electronic states in two-dimensional cuprates. However, an effective one-band Hubbard obtained from the extended three-band Hubbard model containing Cu-O nearest-neighbor repulsion has nearest-neighbor Coulomb interactions~\cite{Simon96,Simon97}. Therefore, it is important to investigate the effect of the nearest-neighbor Coulomb interactions on electronic states in cuprates. For example, screening effect for the on-site Coulomb interaction~\cite{Vandenbrink96,Vandenbrink97,Ayral13,Terletska17} as well as charge-density-wave state~\cite{Zhang89,Yan93,Ohta94} due to such non-local Coulomb interactions have been investigated. In a one-dimensional organic Mott insulator, the emergence of pump-pulse-induced biexcitons has recently been suggested as a consequence of the effect of nonlocal Coulomb interactions~\cite{Kakizaki}. In two-dimensional Mott insulators, the formation of biexcitons has been suggested theoretically~\cite{Gomi2005}. However, there is no theoretical investigation on the effect of the nearest-neighbor Coulomb interaction on pump-probe optical spectroscopy.

In this paper, we theoretically investigate pump-probe optical responses for the single-band extended Hubbard model describing cuprates. The time-dependent optical conductivity just after pumping is calculated by using a time-dependent Lanczos-type exact diagonalization method. At half filling, two types of photoinduced absorptions emerge inside the Mott gap after pumping, being consistent with the experiments~\cite{Okamoto11}. One is attributed to low-energy excitations due to the independent motion of photoinduced holons and doublons, by comparing with optical conductivity for systems with one holon or one doublon. The other is excitations just below the Mott gap that are enhanced by decreasing the on-site Coulomb interaction. The nearest-neighbor (NN) Coulomb interaction controls this enhancement. This is an indication of the presence of a biexciton where two hole-doublon pairs are bounded. The biexcitonic contribution weakens by the introduction of the next-nearest-neighbor (NNN) Coulomb interaction as expected from the spatial distribution of the biexciton. This result clearly indicates that the enhancement of spectral weight below the Mott gap after pumping is not only due to the temperature effect as suggested by the experiment~\cite{Okamoto11, Novelli14} but also the presence of biexcitons. In hole-doped systems, we also find the enhancement of spectral weights after pumping below remnant Mott-gap excitations, which is consistent with the experiment~\cite{Peli17}. The weights are less sensitive to the NN Coulomb interaction but decrease with the NNN Coulomb interaction, indicating the formation of a biexcitonic state in the presence of hole carriers. 

This paper is organized as follows. The extended Hubbard model and time-dependent optical conductivity are introduced in Sec.~\ref{Sec2}. In Sec.~\ref{Sec3}, we calculate the time-dependent optical conductivity at half filling just after pumping. The time-dependent optical conductivity for hole-doped systems is shown in Sec.~\ref{Sec4}. Finally, a summary is given in Sec.~\ref{Sec5}.

\section{Model and Method}
\label{Sec2}
We define an extended version of the Hubbard model in two dimensions as 
\begin{eqnarray}
H&=&-t_\mathrm{h}\sum_{\langle i,j\rangle,\sigma} \left( c^\dagger_{i,\sigma} c_{j,\sigma} + \mathrm{H.c.}\right) -t'_\mathrm{h}\sum_{\langle\langle i,j\rangle\rangle,\sigma} \left( c^\dagger_{i,\sigma} c_{j,\sigma} + \mathrm{H.c.}\right)  \nonumber \\
&&+ U\sum_i n_{i,\uparrow}n_{i,\downarrow}  + V_1\sum_{\langle i,j\rangle} n_i n_j+ V_2\sum_{\langle\langle i,j\rangle\rangle} n_i n_j,
\label{H}
\end{eqnarray}
where $c^\dagger_{i\sigma}$ is the creation operator of an electron with spin $\sigma$ at site $i$, $n_{i,\sigma}=c^\dagger_{i,\sigma}c_{i,\sigma}$, $n_i=\sum_\sigma n_{i,\sigma}$, the summation $\langle  i,j\rangle$ and $\langle \langle i,j\rangle\rangle$ run over pairs of NN and NNN sites, respectively, and $t_\mathrm{h}$, $t_\mathrm{h}'$, $U$, $V_1$, and $V_2$ are the NN hopping, the NNN hopping, the on-site Coulomb interaction, the NN Coulomb interaction, and the NNN Coulomb interaction, respectively. Taking $t_\mathrm{h}$ ($\sim0.4$~eV) to be the unit of energy ($t_\mathrm{h}=1$), we use $V_1=1$, $V_2=0$, and $t'_\mathrm{h}=-0.25$ as a realistic set of parameters for cuprate superconductors~\cite{Tohyama15,Greco17} without being otherwise specified. Since the difference of coordinate numbers around copper influences the value of $U$ in cuprates~\cite{Jang16}, we leave $U$ as a parameter. 

Optical responses obtained by ultrafast pump-probe optical measurements can be described by the time-dependent optical conductivity whose spectral weights are slightly dependent on the shape of probe pulse~\cite{Shao16}. We assume a short probe pulse with a Gaussian form of vector potential. In this case, the diagonal element of the real part of time-dependent optical conductivity after turning off the pump pulse reads~\cite{Filippis12,Shao16}
\begin{eqnarray}
&&\mathrm{Re} \sigma_{\alpha\alpha}(\omega,t) \nonumber \\
&&=\frac{1}{L\omega}\mathrm{Im} \int_0^\infty ie^{i\left(\omega+i\eta\right)s} \langle \Psi(t) \left| [j_\alpha(s),j_\alpha] \right| \Psi(t) \rangle ds  \nonumber \\
&&=\frac{1}{L\omega} \sum_{m,n} \frac{\eta}{(\omega+\varepsilon_m -\varepsilon_n)^2+\eta^2} \nonumber \\
&&\ \ \ \ \ \ \ \ \ \ \ \ \times \big[  \left\langle \Psi(t)  \right| m \rangle \langle m \left| j_\alpha \right| n \rangle \langle n \left| j_\alpha \right| \Psi(t) \rangle \nonumber \\
&&\ \ \ \ \ \ \ \ \ \ \ \ \ \  -  \langle \Psi(t) \left| j_\alpha \right| m \rangle \langle m \left| j_\alpha \right| n \rangle \langle n \left| \Psi(t) \right\rangle \big],
\label{sigmat}
\end{eqnarray}
where $\omega>0$, $L$ is the total number of lattice sites, $\Psi(t)$ is the time-dependent wave function, $\left| m\right\rangle$ is the eigenstates of $H$ with eigenenergy $E_m$, and $\eta$ is a small positive number.  $j_\alpha(s)=e^{iHs}j_\alpha e^{-iHs}$, where $j_\alpha$ is the $\alpha$ component of  current operator given by 
\begin{eqnarray}
j_\alpha&=&it_\mathrm{h}\sum_{\langle i,j\rangle,\sigma} (\mathbf{R}_{ij})_\alpha \left( c^\dagger_{i,\sigma} c_{j,\sigma} - \mathrm{H.c.}\right) \nonumber \\
&&+it'_\mathrm{h}\sum_{\langle\langle i,j\rangle\rangle,\sigma} (\mathbf{R}_{ij})_\alpha \left( c^\dagger_{i,\sigma} c_{j,\sigma} - \mathrm{H.c.}\right)
\label{j}
\end{eqnarray}
with $(\mathbf{R}_{ij})_\alpha$ being the $\alpha$ component of $\mathbf{R}_{ij}=\mathbf{R}_i-\mathbf{R}_j$. 

We use square-lattice periodic Hubbard clusters with $\sqrt{10}\times\sqrt{10}$ and $4\times 4$ sites. We incorporate an external electric field applied along the $x$ direction of the clusters via the Peierls substitution in the hopping terms, $c^\dagger_{i,\sigma} c_{j,\sigma} \rightarrow e^{-iA(t)(\mathbf{R}_{ij})_x} c^\dagger_{i,\sigma} c_{j,\sigma}$. Here $A(t)$
 is the vector potential along the $x$ direction given by
\begin{equation}
A(t)=A_0 e^{-(t-t_0)^2/(2t_\mathrm{d}^2)} \cos[\omega_\mathrm{p}(t-t_0)], 
\label{A}
\end{equation}
where a Gaussian-like envelope around $t_0$ is used with $t_\mathrm{d}$ to characterize the temporal width of pump pulse, and $\omega_\mathrm{p}$ is the central frequency. We set $A_0=0.5$, $t_0=0$, $t_\mathrm{d}=0.5$, and $\omega_\mathrm{p}=U$ without being otherwise specified.

We employ a time-dependent Lanczos method to evaluate $|\Psi(t)\rangle$. The key formula is
\begin{equation}
|\psi(t+\delta{t})\rangle\simeq\sum_{l=1}^{M}{e^{-i\epsilon_l\delta{t}}}|\phi_l\rangle\langle\phi_l|\psi(t)\rangle,
\label{tLancozs}
\end{equation}
where $\varepsilon_l$ and $|\phi_l\rangle$ are eigenvalues and eigenvectors of the tridiagonal matrix generated in the Lanczos iteration, respectively, $M$ is the dimension of the Lanczos basis, and $\delta t$ is the minimum time step. We set $M=50$ and $\delta t=0.02$.

\section{Half filling}
\label{Sec3}

\begin{figure}[t]
\includegraphics[width=0.45\textwidth]{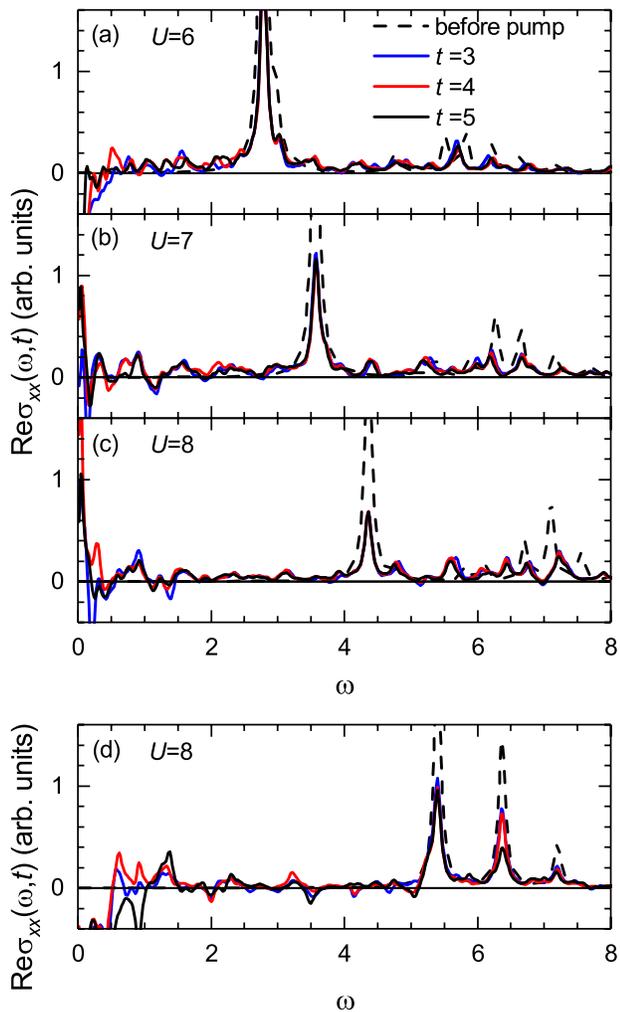}
\caption{The $x$ component of time-dependent optical conductivity after pumping in the half-filled $4\times 4$ extended Hubbard lattice with $t_\mathrm{h}=1$, $t'_\mathrm{h}=-0.25$, and $V_1=1$. (a) $U=6$, (b) $U=7$, and (c) $U=8$. (d) The same as (c) but for the $\sqrt{10}\times\sqrt{10}$ lattice. The broken line represents the optical conductivity before pumping. The blue, red, and black lines represent $\mathrm{Re}\sigma_{xx}(\omega,t)$ at $t=3$, 4, and 5, respectively, with $\eta=1/16$.}
\label{fig1}
\end{figure}

In Figs.~\ref{fig1}(a)-\ref{fig1}(c), we show $\mathrm{Re} \sigma_{xx}(\omega,t)$ in the half-filled $4\times 4$ extended Hubbard model for $U=6,\ 7$, and 8, respectively.  In all cases, a Mott-gap absorption peak before pumping, for example, at $\omega=2.8$ for $U=6$ in Fig.~\ref{fig1}(a), loses its spectral weight after pumping. In addition, there appear many fine structures inside the Mott gap. This is common to a smaller lattice with $\sqrt{10}\times\sqrt{10}$ sites as shown in Fig.~\ref{fig1}(d) for $U=8$. Since there is no prominent feature in the fine structures, it is not easy to identify the origin of them. We thus approach the origin by investigating the same model but with an unrealistic parameter set for cuprates.

\begin{figure}[t]
\includegraphics[width=0.45\textwidth]{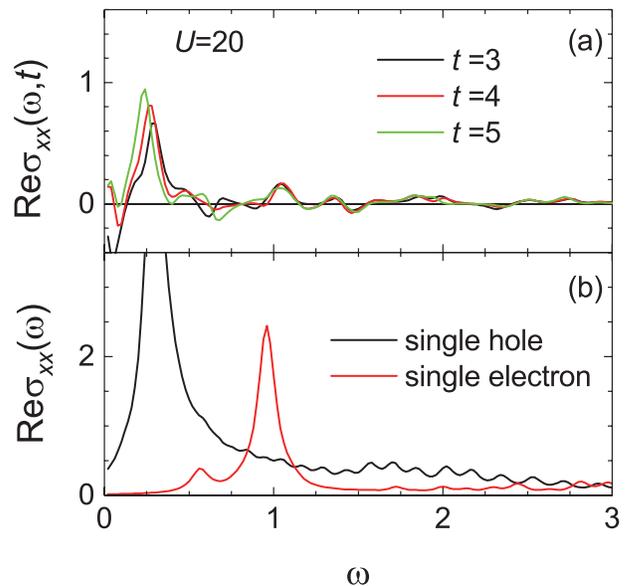}
\caption{(a) $\mathrm{Re}\sigma_{xx}(\omega,t)$ in the half-filled $4\times 4$ extended Hubbard lattice with $t_\mathrm{h}=1$, $t'_\mathrm{h}=-0.25$, and $U=20$. The blue, red, and black lines represent data at $t=3$, 4, and 5, respectively. (b) Equilibrium optical conductivity $\mathrm{Re}\sigma_{xx}(\omega)$ in the $4\times 4$ extended Hubbard lattice with single hole (black line) and single electron (red line). The parameters are the same as (a).}
\label{fig2}
\end{figure}

We first focus on low-energy structures below $\omega\sim2$, for example, at $\omega\sim 0.8$ and 1.7 for $U=8$ [see Fig.~\ref{fig1}(c)]. Pump photon generates holons and doublons. Therefore, the two particles may propagate incoherently in the spin background with finite energy. This will be a possible origin of the low-energy structures. This is evident from the case of large $U$. Figure~\ref{fig2}(a) shows $\mathrm{Re}\sigma_{xx}(\omega,t)$ for $U=20$. There are nearly time-independent peak structures at $\omega\sim0.3$ and 1.1. The energy position of the former corresponds to the peak position of equilibrium optical conductivity $\mathrm{Re}\sigma_{xx}(\omega)$ for single-hole motion given by the $4\times 4$ cluster with 15 electrons, as seen in Fig.~\ref{fig2}(b). The latter is close to the peak position of $\mathrm{Re}\sigma_{xx}(\omega)$ for single-electron motion.  We note that the difference of the peak energies between the single-hole and single-electron motions is due to the effect of $t'_\mathrm{h}$ that induces asymmetry between hole and electron doping in terms of antiferromagnetic correlation~\cite{Tohyama01}.  The good correspondence between the peak positions in Figs.~\ref{fig2}(a) and \ref{fig2}(b) strengthens the speculation that incoherent motion of single holon and single doublon emerges in $\mathrm{Re}\sigma_{xx}(\omega,t)$ even for a realistic value of $U$. Such an assignment of the two structures has been done in the pump-probe experiments for La$_2$CuO$_4$ and Nd$_2$CuO$_4$~\cite{Okamoto11}.

We next discuss the photoinduced structures above $\omega\sim2$ in Fig.~\ref{fig1}. We find a slight increase of spectral weight below the absorption peak with decreasing $U$ from $U=8$ to $U=6$. In order to obtain a crucial insight for this $U$ dependence, we adopt the following procedure.

Since probe photons excite the time-evolving state after pumping, it is natural to investigate photo-excitations from the most dominant configuration in the state. Taking into account the formation of a holon-doublon bound pair and its large contribution to the absorption peak, we assume that the state giving the largest absorption peak contributes predominantly to the time-evolving state. We therefore regard an eigenstate forming the absorption peak at the Mott gap, $\left|\phi_\mathrm{peak}\right>$, as the initial state before applying a probe photon~\cite{Takahashi02}. Calculating $W_m=\left| \left< m\right| j_x\left|\phi_\mathrm{peak}\right> \right|^2$ and integrating it over an energy interval $\Delta$ corresponding to the energy region lower than the absorption-peak energy $\varepsilon_\mathrm{peak}$, i.e.,
\begin{equation}
I(\Delta)=\int_{\varepsilon_\mathrm{peak}-\Delta}^{\varepsilon_\mathrm{peak}} d\omega \sum_m W_m \delta \left( \omega-\varepsilon_m+\varepsilon_\mathrm{peak} \right),
\label{Int}
\end{equation}
we are able to discuss the $U$ dependent spectral weight in $\mathrm{Re}\sigma_{xx}(\omega,t)$ more quantitatively.

\begin{figure}[t]
\includegraphics[width=0.45\textwidth]{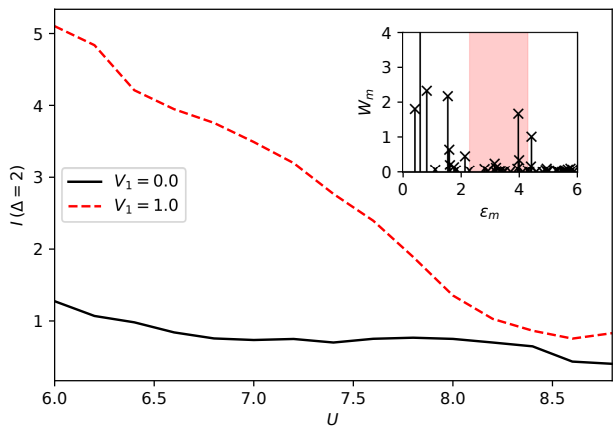}
\caption{Integrated weight $I(\Delta)$ at $\Delta=2$ (see text) as a function of $U$ in the half-filled $\sqrt{10}\times \sqrt{10}$ extended Hubbard lattice with $t_\mathrm{h}=1$ and $t'_\mathrm{h}=-0.25$. $V_1=0$ (solid black line) and $V_1=1$ (dashed red line). Inset shows $W_m$ for $V_1=1$ and $U=6.6$ and the colored region corresponds to the energy interval $\Delta$. The bar with cross denotes $W_m$ for corresponding $\varepsilon_m$.}
\label{fig3}
\end{figure}

For this purpose, we use a $\sqrt{10}\times\sqrt{10}$ periodic lattice of the same model, since all eigenstates including $\left|\phi_\mathrm{peak}\right>$ are numerically obtained.  We note that the $4\times 4$ and $\sqrt{10}\times\sqrt{10}$ clusters give similar spectral properties of $\mathrm{Re}\sigma_{xx}(\omega,t)$. Figure~\ref{fig3} exhibits the $U$ dependence of the integrated wight $I(\Delta)$ at $\Delta=2$. The colored region in the inset shows the integrated energy region, where contributions from the propagation of holons and doublons are expected to be small. For $V_1=1$ the integrated weight increases with decreasing $U$ as is expected from Fig.~\ref{fig1}. For $V_1=0$, however, such an increase is small as compared with that for $V_1=1$. This clearly indicates that the increase of the weight below the absorption peak comes from $V_1$.

\begin{figure}[t]
\includegraphics[width=0.45\textwidth]{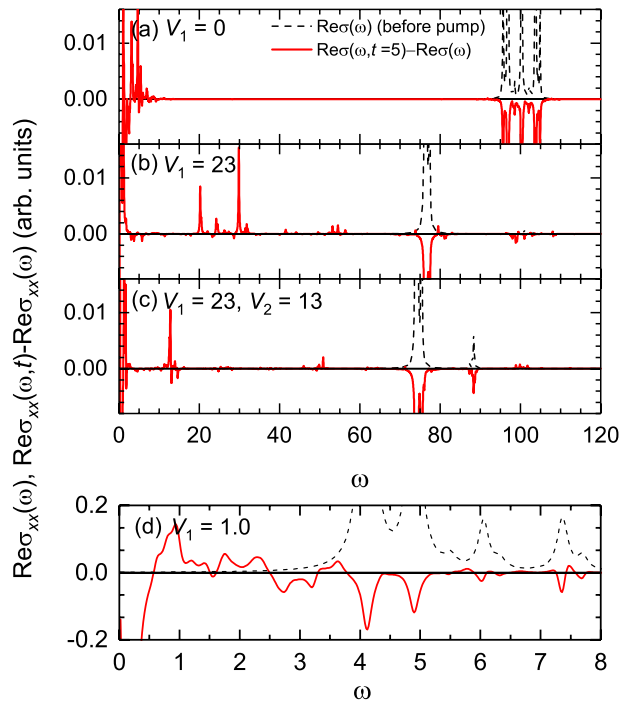}
\caption{The $x$ component of time-dependent optical conductivity in the half-filled $\sqrt{10}\times \sqrt{10}$ extended Hubbard lattice with $t_\mathrm{h}=1$, $t'_\mathrm{h}=-0.25$, and $U=100$. (a) $V_1=0$, (b) $V_1=23$, and (c) $V_1=23$ and $V_2=13$. (d) $U=6$ and $V_1=1$. The broken line represents $\mathrm{Re}\sigma_{xx}(\omega)$ before pumping. The red line represents $\mathrm{Re}\sigma_{xx}(\omega,t)-\mathrm{Re}\sigma_{xx}(\omega)$ at $t=5$. $A_0=8$, $\omega_\mathrm{p}=U$, and $\eta=0.1$.}
\label{fig4}
\end{figure}

In order to clarify the effect of $V_1$ on pump-probe optical conductivity, we further investigate $\mathrm{Re}\sigma_{xx}(\omega,t)$ for large $U$ in the $\sqrt{10}\times \sqrt{10}$ lattice. Making $U$ large, we can separate the effect of $V_1$ from others coming from the motion of holons and doublons. Figure~\ref{fig4} shows $\mathrm{Re}\sigma_{xx}(\omega,t)$ at $t=5$ for $U=100$. Before pumping, $\mathrm{Re}\sigma_{xx}(\omega)$ (black dotted lines) exhibit a single-exciton behavior for large $V_1$, which is seen, for example, as a peak at $\omega=76$ separated from continuum at around $\omega=100$ in Fig.~\ref{fig4}(b). Without $V_1$, there is no photoinduced spectral weight except for $\omega<10$ [Fig.~\ref{fig4}(a)]. In contrast, photoinduced spectral weights emerge when $V_1$ is finite as shown in Fig.~\ref{fig4}(b) ($V_1=23$). There is a large peak at $\omega=30$, whose position is low in energy by approximately $2V_1=46$ from the absorption peak at $\omega=77$. A pump photon creates a holon-doublon pair and a subsequent probe photon creates another pair. If the two pairs bind, one expects a bound state called biexciton in the final state of the pump-probe process. In fact, counting the energy for the immobile biexciton, where the two doublons are located at the NNN sites, i.e., diagonal positions, we obtain the energy $2U-4V_1$, which is lower than the energy for two separated holon-doublon pairs ($2U-2V_1$) by $2V_1$. This is an interpretation of the peak at $\omega=30$. This interpretation is confirmed by introducing the NNN Coulomb interaction $V_2$, i.e., diagonal repulsive interaction, which prevents diagonal positions of two doublons. By $V_2=13$, the immobile biexciton increases its energy to $2U-4V_1+4V_2=160$, which is larger than $2U-2V_1=154$, indicating the absence of a biexciton. In fact, there is no bound state in Fig.~\ref{fig4}(c). We note that another peak at $\omega=20$ in Fig.~\ref{fig4}(b) as well as that at $\omega=12$ in Fig.~\ref{fig4}(c) is due to the breaking of a holon-doublon pair by probe photon. 

We note such biexcitonic states are also seen for a more realistic value of $U$. Figure~\ref{fig4}(d) shows the case with $U=6$ and $V_1=1$, where small peaks appear at around $\omega=2$ lower in energy by $2V_1$ than the lowest absorption peak at $\omega=4.2$.

A remaining question is why the presence of $V_1$ enhances the $U$ dependence of $I(\Delta)$ in Fig.~\ref{fig3}. This may be related to a crucial role of spin background on biexcitonic states. With decreasing $U$, effective antiferromagnetic (AF) interaction $J$ between spins in the background increases because of $J\sim4t_h^2/U$. The AF background assists the formation of biexcitonic states to gain magnetic energy. This might be the reason for the increase of $I(\Delta)$ even for $V_1=0$ in Fig.~\ref{fig3}. Introducing $V_1$ into such an AF background will induce cooperative effects to strengthen biexciton formation, leading to the enhancement of $U$ dependence. If the $U$ dependence is realized in real cuprate materials, we expect that photoinduced spectral weight below the Mott gap is larger in Nd$_2$CuO$_4$ than in La$_2$CuO$_4$ because of smaller $U$ in Nd$_2$CuO$_4$ than in La$_2$CuO$_4$~\cite{Jang16}.

\section{Hole doping}
\label{Sec4}

\begin{figure}[t]
\includegraphics[width=0.45\textwidth]{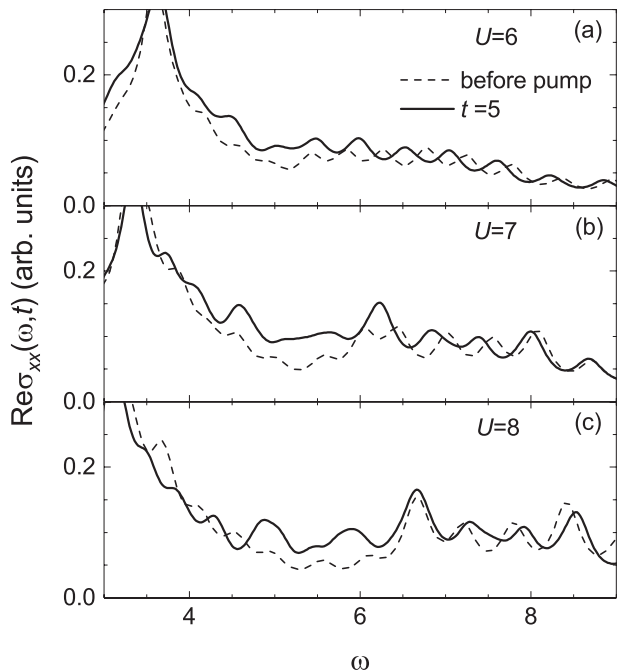}
\caption{The $x$ component of time-dependent optical conductivity in the two-hole doped $4\times 4$ extended Hubbard lattice with $t_\mathrm{h}=1$, $t'_\mathrm{h}=-0.25$, and $V_1=1$. (a) $U=6$, (b) $U=7$, and (c) $U=8$. The broken line represents the optical conductivity before pumping. The solid line represents $\mathrm{Re}\sigma_{xx}(\omega,t)$ at $t=5$.}
\label{fig5}
\end{figure}

Carriers introduced into Mott insulators change the electronic states significantly. Not only metallic behavior but also the reconstruction of excited states across the Mott gap appears~\cite{Uchida91,Dagotto92,Tohyama05}. The broken lines in Fig.~\ref{fig5} exhibit $\mathrm{Re}\sigma_{xx}(\omega)$ for a doped $4\times 4$ extended Hubbard lattice with hole concentration $x=2/16=0.125$, The peak  seen at the Mott-gap energy (see Fig.~\ref{fig1}) is washed out and a remnant peak is seen, for instance, at $\omega\sim 6.6$ for $U=8$ [see Fig.~\ref{fig5}(c)]. It is known that the remnant peak is higher in energy than the original Mott-gap peak~\cite{Uchida91,Dagotto92,Tohyama05}. We also note that the low-energy excitations below $\omega=4$ are metallic components but with finite excitation energy. 

Pumping such a doped system induces little change in optical conductivity as shown by the solid lines that are $\mathrm{Re}\sigma_{xx}(\omega,t)$ at $t=5$ in Fig.~\ref{fig5}. Nevertheless, one can identify systematic enhancement of the spectral weight below the remnant Mott gap, which is seen in the energy region of $4.7<\omega<6.5$ for $U=8$, $4.4<\omega<6$ for $U=7$, and $4<\omega<5.5$ for $U=6$. This enhancement is less sensitive to $V_1$ (not shown).

\begin{figure}[t]
\includegraphics[width=0.45\textwidth]{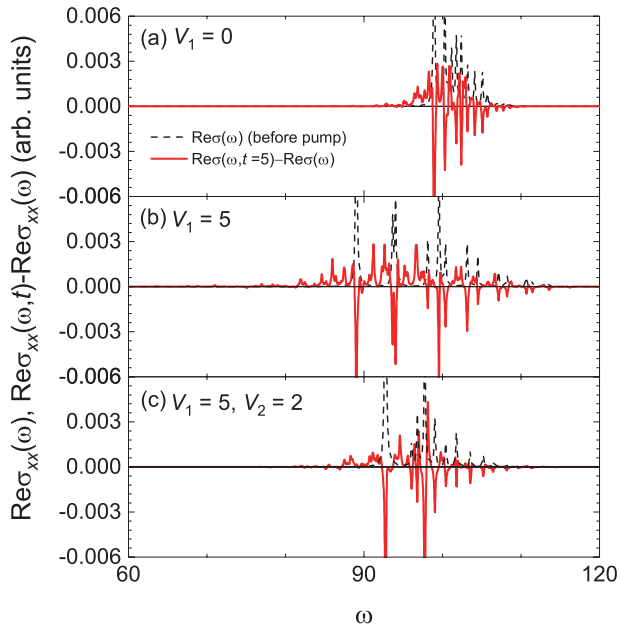}
\caption{The $x$ component of time-dependent optical conductivity in the two-hole doped $\sqrt{10}\times \sqrt{10}$ extended Hubbard lattice with $t_\mathrm{h}=1$, $t'_\mathrm{h}=0$, and $U=100$. (a) $V_1=0$, (b) $V_1=5$, and (c) $V_1=5$ and $V_2=2$. The black broken line represents $\mathrm{Re}\sigma_{xx}(\omega)$ before pumping. The red line represents $\mathrm{Re}\sigma_{xx}(\omega,t)-\mathrm{Re}\sigma_{xx}(\omega)$ at $t=5$. $A_0=8$, $\omega_\mathrm{p}=U$, and $\eta=0.2$.}
\label{fig6}
\end{figure}

In order to clarify this enhancement, we again examine the large $U$ case for the $\sqrt{10}\times \sqrt{10}$ cluster. Figure~\ref{fig6} shows $\mathrm{Re}\sigma_{xx}(\omega)$ (black broken lines) and difference spectra $\mathrm{Re}\sigma_{xx}(\omega,t=5)-\mathrm{Re}\sigma_{xx}(\omega)$ (red solid lines) for $U=100$.  Without $V_1$, we clearly find the increase of spectral weights below remnant Mott-gap excitations near $\omega=100$ as shown in Fig.~\ref{fig6}(a). This is in contrast to the half-filled case where there is no induced spectral weight just below the Mott gap [see Fig.~\ref{fig4}(a)]. Induced spectral weights are also present at $V_1=5$ [Fig.~\ref{fig6}(b)] with an energy scale of $V_1$ below the lowest-energy Mott-gap excitation, but there is no dramatic change as compared with the case of $V_1=0$. In the half-filled case, $V_1$ is crucial for the enhancement of spectral weight originating from the formation of biexcitonic states. Therefore, we suppose that such a biexcitonic state will be formed in the hole-doped system with the help of hole carriers. Here, this view reminds us of the formation of charged stripe order in the doped Mott
insulator~\cite{Zaanen89, Prelovsek93, Prelovsek94, Nayak97, White98, Zaanen98, Zheng17}. To obtain the gain of hopping and magnetic energy, holes in the Mott insulator construct stripe configuration. If a site between the diagonal stripes is doubly occupied, the doublon site is surrounded by the holes, resulting in an energy gain due to virtual electron hoppings from doublon to hole. If this is the case, the introduction of $V_2$ will reduce spectral weight because of the suppression of the biexcitonic state. In fact, $V_2$ reduces the spectral weight as shown in Fig.~\ref{fig6}(c). This indirectly justifies the formation of a biexcitonic state.

In contrast, such increase of spectral weights below remnant Mott-gap excitations is not found in the one-dimensional hole-doped system for $U=100$ in a ten periodic lattice (not shown). In the one-dimensional system, hole binding is not relevant due to spin-charge separation, and holes are not efficiently collected around photoinduced doublons, yielding less energy gain due to virtual hopping between the doublon and hole. Even if biexcitonlike charge configuration is constructed, the energy gain due to virtual electron hopping is less than that in the two-dimensional system, since the number of hopping paths around the doublon is two in one dimension, but four in two dimensions. The energy gain due to the virtual hopping process around the doublon site is closely related to the formation of biexcitons.

Another possible origin of the enhancement of spectral weight after pumping might be related to the propagation of photoinduced holons and doublons in the presence of hole carriers. This comes from the observation that the holons and doublons may act as carriers in addition to the original hole carriers, resulting in the increased spectral weight in optical conductivity~\cite{Tohyama05}. However, such an increase occurs in the energy range of particle-hole excitations in optical conductivity and thus not near Mott-gap excitations~\cite{Tohyama05}. Therefore, this mechanism is not the origin of the enhancement.

The enhancement below the remnant Mott-gap excitations in Fig.~\ref{fig5} is tiny but can be detectable. In fact, a recent pump-probe experiment has reported the emergence of spectral weight~\cite{Peli17}. In the experiment, the enhanced spectral weight below the Mott gap, more precisely charge-transfer (CT) gap, is accompanied by the reduction of the weight at the CT peak, which has been interpreted as a consequence of the redshift of the CT peak based on atomic-level analyses~\cite{Peli17}. In our calculations, the spectral weight reduction of the remnant Mott-gap excitations is unclear in Fig.~\ref{fig5}. However, the reduction is clearly seen in Fig.~\ref{fig6} and thus the amount of reduction may depend on several conditions such as pump-photon energy. Therefore, our results may give an alternative interpretation on the enhancement of spectral weight seen in the experiment.

\section{Summary}
\label{Sec5}

We have investigated pump-probe optical responses for two-dimensional Mott insulators and hole-doped systems by calculating the time-dependent optical conductivity in the single-band extended Hubbard model. At half filling, two types of photoinduced absorptions emerge inside the Mott gap after pumping, consistent with the experiments~\cite{Okamoto11}. One is attributed to low-energy excitations due to the independent propagation of photoinduced holons and doublons. The other is excitations just below the Mott gap that are enhanced by the nearest-neighbor Coulomb interaction. This is an indication of the presence of a biexcitonic state where two doublons are located at the next-nearest-neighbor sites. This is evidenced by the destruction of the biexciton by the second-nearest-neighbor Coulomb interaction. This result clearly indicates that the enhancement of spectral weight just below the Mott gap after pumping is not only due to the temperature effect as suggested by the experiment~\cite{Okamoto11,Novelli14} but also the presence of the biexciton. 

In hole-doped systems, we have found the enhancement of spectral weights after pumping below remnant Mott-gap excitations, which is consistent with the experiment~\cite{Peli17}. The weights are less sensitive to the NN Coulomb interaction but decrease with introducing the NNN Coulomb interaction, indicating the formation of a biexcitonic state in the presence of hole carriers. These findings are important for a full understanding of pump-probe optical responses in cuprate superconductors.

These conclusions have been obtained from small clusters with $4\times 4$ and $\sqrt{10}\times\sqrt{10}$ sites. We believe that the effects of biexcitons extending minimally over four sites can be expressed by such clusters, but it would be desirable to examine the biexcitonic effects using much larger clusters. In addition, in our calculations we do not include the effect of phonon degree of freedom emerging through electron-phonon interactions. It has been reported that an ultrafast reaction of the bosonic field affects the time-dependent optical responses~\cite{Novelli14}. It would be interesting to examine the influence of phonons on the biexcitonic structures. Therefore, these two unexplored issues remain as future problems.

\begin{acknowledgments}
We would like to thank H. Okamoto for fruitful discussions. This work was supported by the Japan Society for the Promotion of Science, KAKENHI (Grant No. 26287079), by CREST (Grant No. JPMJCR1661), Japan Science and Technology Agency, by the creation of new functional devices and high-performance materials to support next-generation industries (GCDMSI) to be tackled by using post-K computer, and by MEXT HPCI Strategic Programs for Innovative Research (SPIRE) (hp160222 and hp170274).
\end{acknowledgments}

\nocite{*}


\end{document}